\documentstyle[epsf,11pt,psfig]{article}
\input{psfig}
\renewcommand{\theequation}{\arabic{section}.\arabic{equation}}
\begin{document}
\title{ \bf{ On Zero Modes and the Vacuum Problem -- \\
A Study of Scalar Adjoint Matter in Two-Dimensional Yang-Mills Theory
        via Light-Cone Quantisation.} }
\author{ {\it Alex C. Kalloniatis}\\
Max-Planck-Institut f\"ur Kernphysik \\
D-69029 Heidelberg  \\}
\date{15 August, 1995.}
\maketitle

\begin{abstract}
SU(2) Yang-Mills Theory coupled to massive adjoint scalar matter is
studied in (1+1) dimensions using Discretised Light-Cone Quantisation.
This theory can be obtained from
pure Yang-Mills in 2+1 dimensions via dimensional reduction.
On the light-cone, the vacuum structure of this theory is encoded
in the dynamical zero mode of a gluon and a constrained mode of
the scalar field.
The latter satisfies a linear
constraint, suggesting no nontrivial vacua in the present
paradigm for symmetry breaking on the light-cone.
I develop a diagrammatic method to solve the constraint
equation. In the adiabatic approximation I compute the
quantum mechanical potential governing the
dynamical gauge mode. Due to a condensation of the lowest
momentum modes of the dynamical gluons, a centrifugal barrier
is generated in the adiabatic potential. In the present theory
however, the barrier height appears too small to make any impact in this
model. Although the theory is superrenormalisable on naive
powercounting grounds, the removal of ultraviolet divergences
is nontrivial when the constrained mode is taken into account.
The open aspects of this problem are discussed in detail.
\end{abstract}
Preprint: MPI-H-V29-1995

\newpage
\section{Introduction}
For some years now intensive work has gone into
developing light-cone field theoretic methods
\cite{Wei66,LeB80,PaB85,PHW90} for
solving the problem of the hadron spectrum from
Quantum Chromodynamics. The fundamental advantage of
Dirac's `front form' Hamiltonian framework \cite{Dir49} is the
simple vacuum structure and the (related) positivity
of the momentum operator. With these one can foresee
a picture of hadrons emerging from a diagonalisation of
the light-cone Hamiltonian consistent within the intuitive
picture of the constituent quark model (see also \cite{HeK95}).

However, two problems have been the stumbling block to
this program: renormalisation and the vacuum problem.
My main concern in the present work is the
latter problem. Ironically, for all the advantages of a simple vacuum
for understanding hadron structure, the sophisticated
field theoretic picture of QCD as a gauge field theory
demands {\it some} non-trivial vacuum structure associated with
the non-Abelian group topology. In other words, the
constituent quark picture works well, but it doesn't
work {\it everywhere} in strong-interactions. A reasonable
`solution' to QCD must encompass both features.

A scenario in which vacuum structure and the advantages
of the front-form could sensibly
co-exist was recently put forward by Robertson \cite{Rob93}.
In the infra-red regularisation achieved by
Discretised Light-Cone Quantisation (DLCQ) of a given
field theory, some zero mode field operators are not
dynamical field quanta, thus preserving vacuum simplicity,
but rather satisfy constraint equations \cite{MaY76}.
In some simpler models in which symmetries are known to be
spontaneously broken these constraints happen to possess multiple
solutions. The light-cone symmetry breaking paradigm
thus associates each of these solutions with
a particular vacuum choice in the instant-form treatment
of the same theory.
Thus for example,
the scalar $\phi^4$ model in 1+1 dimensions,
also studied by \cite{HKW91}, has a
zero mode satisfying a cubic equation. One of the solutions
corresponds to the trivial (unbroken phase) vacuum.
The other two reflect the broken phase of the theory.
Thus, for example, one could do perturbation theory
about the classical solution in either phase \cite{Rob93}.
Alternately, van de Sande et al. \cite{BPV93,PiV94,PVH95}
thoroughly show how the Tamm-Dancoff method
can be brought to bear on this problem
with a determination of
condensates and the critical coupling.
Another approach using the $1/N$ expansion for analogous
models has been developed by \cite{BGW95} with similar physics
emerging.

Turning to QCD, the paradigm appears difficult to apply.
The phenomena one seeks are: 1) $\theta$-vacua or their analogue,
2) spontaneous (anomalous) breaking of chiral (scale)
symmetry leading to the appearance of condensates in the
nonperturbative vacuum $|\Omega \rangle$ such as
${\frac{\alpha_s}{\pi}}
\langle \Omega | G^a_{\mu \nu} G^{a \mu \nu} | \Omega \rangle \sim  10^{-2}
\;.$
The value ascribed here leads to a parametrisation of hadronic
resonances via QCD Sum Rules \cite{SVZ79}. All these effects are related to
non-trivial gluonic configurations in the ground state.
If the above paradigm is to be relevant, then constrained zero
modes of the gluons should play a role. Such objects indeed occur
in non-Abelian gauge theory in (3+1) dimensions in DLCQ as shown in
the remarkable paper of Franke et al. \cite{FNP81}.
Working with SU(2) pure glue theory,
they introduced a modified light-cone gauge
consistent with the space compactification,
\begin{equation}
\partial_- A^+ = 0,
\label{maxCoulgauge}
\end{equation}
with an additional colour rotation rendering the zero mode of $A^+$
diagonal in colour space.
It turns out that
a constrained zero mode appears only in the corresponding
diagonal component of the transverse gluons.
There is
only one constraint equation
for the zero mode of the field $A^3_\perp$. This equation
turns out to be linear and thus there is a unique solution \cite{FNP81}.
How `multiple vacua' now emerge in light of this is not clear.

The additional feature in gauge theory
is the presence of an additional, dynamical, zero mode.
In the case of Franke et al., it manifests itself as the surviving
piece of the $A^+$ gauge potential. Dynamical zero modes
{\it in general} mean the naive argument for vacuum triviality
in the front form does not apply. The key questions are: just
{\it how} nontrivial is the vacuum, can one bring it
under control, and do the zero modes produce the
desired vacuum structure?

The context in which I will try to answer these questions
is that of SU(2) Yang-Mills theory coupled to scalar adjoint
matter in (1+1) dimensions. As shown elsewhere, this theory
can be regarded as the dimensional reduction of pure gauge theory
in (2+1) dimensions \cite{DKB93}.
The zero mode structure of this theory was described in
an earlier work \cite{PKP95} wherein it was shown that, in
the same gauge as Franke et al., there is one dynamical and one
constrained zero mode. The spectrum of the theory
has been further studied in \cite{PaB95}.
The approach taken in \cite{PKP95} to the dynamical
mode was similar to that of, for example, Manton \cite{Man85}
in the Schwinger model or of Lenz et al. \cite{LST94} for
the case of QCD(1+1) with
adjoint fermions. The latter is based on a rather new
method of gauge-fixing in the Weyl gauge \cite{LNT94}.
What was lacking in \cite{PKP95} was a
solution to the constraint equation, which the present work
sets out to complete. In this respect, this work is very much
a part in the series of works \cite{KaP94,KaR94}.

The approach for the dynamical, or `gauge', mode in the
above works is in the spirit of the adiabatic approximation.
The gauge mode part of the Hamiltonian is not in fact separable
from the Fock part. Nonetheless,
the gauge mode is `frozen' and the vacuum of the particle sector
determined. The dynamics of the gauge mode is then solved for
in the vacuum of the particle sector. In this theory, there
is no $Z$-vacuum degeneracy. Rather, one expects
a two-fold $Z_2$ degeneracy as evident from the global
symmetries of the `adiabatic potential'
which controls the dynamics of the gauge zero mode.
It is not clear from the literature whether one should
expect any condensate phenomena in this specific model.

In the previous work \cite{PKP95}, this potential was computed
in the absence of the constrained zero mode. It was found to
contain logarithmic divergences for which the counterterms, at
the time, were not evident. However, the result for
the cutoff independent part was of a flat approximately
square well structure. This is rather unlike the result
with adjoint fermions \cite{LST94} where a centrifugal
barrier appears in the adiabatic potential.
It was already noted for scalar matter \cite{PKP95}, that
the absence of fermions and transverse momenta
meant insufficient degrees of freedom
for some composite operator to acquire a condensate.
However, it is at least desirable to see whether there is
sufficient structure in the gluonic ground state to provide the
`seeding' for such non-triviality when further modes are included.

In the present work I therefore address the two main problems which, in
\cite{PKP95}, remained untreated: solving
the constraint equation and the renormalisation problems in
the potential. For the former, I develop
a diagrammatic representation which will guide
a solution of the constraint. The most significant aspect of the solution
is an iteration to all orders of two-point insertions in the
constraint solution.
For the second problem, I am able to report partial success
but difficulties remain. In particular, in the absence of
the constrained zero mode the divergences in the potential
can be naturally handled with mass renormalisation.
Finally, I study the inclusion of the constraint
in the potential. There remain logarithmic divergences which in
the present work are subtracted by hand. In the renormalised potential
one observes that a condensation of the lowest momentum
mode of the scalar field, achieved by the sum to all orders
of the insertion in the constraint solution, leads indeed to the
appearance of a distinct, albeit small, potential barrier.
The barrier is highest in the case of the renormalised mass of
the `gluons' being zero.
If an exact $Z_2$ degeneracy is to arise in
the continuum limit, this barrier would have to
become impenetrable. This does not appear to happen here.
In the final section, I discuss in some detail the open problem
of the remaining logarithmic
divergences. There is a brief summary at the end.
Notation and other details are relegated to the Appendices.

\section{Review of Scalar Adjoint Matter and SU(2) Yang-Mills}
The model-theory I work with is (1+1) dimensional
non-Abelian gauge theory covariantly coupled to massive scalar adjoint matter
\begin {equation}
    {\cal L}  = {\rm Tr} \Bigl(
        - {1\over2} {\bf F}^{\alpha\beta} {\bf F}_{\alpha \beta}
       + {\bf D}^\alpha \Phi   {\bf D}_\alpha \Phi
        - \mu_0^2 \Phi^2 \Bigr)
       \ .
\end   {equation}
The field strength tensor and
covariant derivative $ {\bf D}_\alpha$ are respectively defined
by
\begin{equation}
{\bf D}^\alpha = \partial^\alpha + i g [{\bf A}^\alpha, \cdot ]
\quad {\rm {and}} \quad
{\bf F}^{\alpha \beta} = \partial^\alpha {\bf A}^\beta
  - \partial^\beta {\bf A}^\alpha + i g [{\bf A}^\alpha,
{\bf A}^\beta ]
\;.
\end{equation}
As in \cite{PKP95}, I represent field matrices
in a colour helicity basis:
\begin {equation}
  \Phi = \tau ^3 \varphi_3 + \tau ^+ \varphi_+ + \tau ^- \varphi_-
\; .
\end    {equation}

Treatments of this theory in DLCQ have been given in Refs.
\cite{DKB93,PKP95}.
The equations of motion
are
\begin {equation}
   {\bf D}_\beta  {\bf F}^{\beta \alpha} =  g  {\bf J}^\alpha
   \ , \ {\rm with}\quad
   {\bf J}^\alpha = - i \bigl[ \Phi , {\bf D}^\alpha \Phi \bigr]
 \ ,\quad\  {\rm and} \quad
   ( {\bf D}^\alpha  {\bf D}_\alpha + \mu_0^2) \Phi  = 0 \ .
\label{eqofmot}
\end    {equation}
Note that the `matter current' $ {\bf J}^\alpha$ is not conserved,
$\partial_\alpha {\bf J}^\alpha \neq 0$,
whereas the
total `gluon current'
$ {\bf J}^\alpha_{\rm G} =   {\bf J}^\alpha
  -i\bigl[ {\bf F}^{\alpha\beta} , {\bf A}_\beta \bigr] $
is conserved.
I use the light-cone Coulomb gauge
$ \partial_-  {\bf A}^+ = 0$, which preserves the zero mode
of ${\bf A^+}$.
Then a single rotation in colour space
suffices to diagonalise the SU(2) colour matrix $ {\bf A}^+ = A^+_3 \tau_3$.
Finally, the diagonal zero mode of $ {\bf A}^-(x^+_0)$ can be gauged
away \cite{KPP94} at some fixed light-cone time $x^+_0$.
For writing the Hamiltonian, it is convenient to choose
this time as $x^+_0 = 0$, the null-plane initial value surface on which we
specify the independent fields.
The quantum mode $ A^+_3 $ has a conjugate momentum
$  p \equiv \delta L / \delta  (A^+_3)  = 2 L \partial_+ A^+_3  $
and satisfies the commutation relation
\begin {equation}
 \bigl[ A^+_3 , p \bigr] =
 \bigl[ A^+_3 , \ 2 L \partial_+ A^+_3 \bigr] = i \ .
\end    {equation}
As in \cite{PKP95}, I shall work in Schr\"odinger representation
for this quantum mechanical degree of freedom.
In the following it will be useful to invoke the dimensionless
combination
\begin {equation}
   z \equiv {{g  A^+_3  L} \over \pi} \ .
\end {equation}
There are additional global symmetries which can be seen in
terms of this mode. First,
Gribov copies \cite{Gri78,Sin78} correspond to
shifts $ z \rightarrow  z + 2n, n \in Z$,
see Appendix B.
Shifts $z \rightarrow z + (2n + 1), n\in Z$ are `copies' generated by
the group of centre conjugations of SU(2),
namely $Z_2$ symmetry \cite{Lus83}.
The finite interval $0 <  z  < 1$ is called the
fundamental modular domain, see for example \cite{vBa92}.
Two further symmetries are important: Weyl reflection,
$z \rightarrow - z$ and the composition of a reflection
and centre transformation leading to a symmetry of the theory
under $z \rightarrow (1 - z)$. After selection of the fundamental
domain, this last is the only remnant symmetry leading to symmetry
under reflection about $z=\frac{1}{2}$.
To further prepare the reader, it will later become convenient
to switch to a variable $\zeta = (z - \frac{1}{2})$ in
the fundamental modular domain.

The diagonal component of the hermitian colour matrix $ \Phi $
is $ \varphi_3$.
The quantisation, with the exception of the zero mode
$ {\hbox{\vbox{\ialign{#\crcr
    ${\,\scriptstyle \circ\,\,}$\crcr
   \noalign{\kern1pt\nointerlineskip}
    $\displaystyle{\varphi}_{3}$\crcr}}}}
=  a_0 / \sqrt{4 \pi} $, is straightforward.
At $x^+=0$, I expand in momentum modes
\begin {equation}
  \varphi_3 (x^-) = {a_0\over \sqrt{4\pi}}
   + {1 \over \sqrt{4\pi}} \sum_{l =1}^{\infty} \Bigl(
  \  a_l \  w_l \ {\rm e}^{-i l   {\pi\over L}   x^-}
  + \  a^{\dag}_l \  w_l \ {\rm e}^{+i l   {\pi\over L}   x^-}
  \Bigr) \ .
\end    {equation}
where  $ w_l  = 1/ \sqrt{l} $.
Note that I will reserve $l,l',l_1, \dots$ for
the nonzero integer valued momenta of the real scalar field.
The momentum field conjugate to $ \varphi_3$ is
$ \pi^3 = \partial_- \varphi_3$.
The quantum commutation relation at equal $ x^+$ for the normal modes is
given in Appendix C. However, it leads to
the Fock commutator
$     \bigl[ a_l , a^{\dag}_{l'} \bigr] = \delta_{l,l'} $
($l,l' > 0 $).
Sometimes it will be convenient to write the Kronecker
$\delta_{l , l'}$ as $\delta_l^{l'} $.
As the zero mode of $\pi_3$ vanishes, the zero mode of $\varphi$
is not a degree of freedom, but will turn out to satisfy a constraint
which is the main point of this paper.

The off-diagonal components of $\Phi$
are complex fields with
$ \varphi_+ (x^-) = \varphi_-^{\dag} (x^-) $.
The momentum conjugate to $ \varphi_-$ is
$ \pi^- = \bigl(\partial_- + i  g  v \bigr) \varphi_+$.
The other conjugate pair is obtained
simply by hermitian conjugation.
A way of quantizing this field which is not complicated by
the large gauge transformations
was discussed in \cite{PKP95}. The expansion is
over {\it half-integer} momenta but in a manner
consistent with periodic boundary conditions, namely
\begin {equation}
  \varphi_- (x) =
    {{\rm e}^{+ i m_0  {\pi\over L}   x^-} \over \sqrt{4\pi}}\ %
  \sum_{m={1\over 2}}^{\infty} \Bigl(
 \  b_m \  u_m \ {\rm e}^{-i  m   {\pi\over L}   x^-}
 + \  d^{\dag}_m \  v_m \ {\rm e}^{+i  m   {\pi\over L}   x^-} \Bigr)
\ . \label{fockexp} \end    {equation}
where $ u_m (z) = 1/\sqrt{m + \zeta}$ and
$ v_m (z) = 1/\sqrt{m - \zeta}$.
The objects $m_0$ and $\zeta$ are functions of $z$,
defined by $m_0(z) = ({\rm{integer \, part \, of\,}} z) - \frac{1}{2},
\zeta(z) = z - m_0(z)$. They
satisfy the relations
\begin {equation}
  m_0 (z+1) = m_0 (z) + 1, \quad m_0  (-z)= -m_0  (z), \quad
\zeta  (z+1) = \zeta  (z), \quad \zeta (-z)= -\zeta (z)
 \ . \label {zsymm} \end   {equation}
The domain interval is now $ - {1\over 2} < \zeta (z) < {1\over 2} $
for all values of $ z $. For the fundamental domain
$m_0 = -\frac{1}{2}$, but the specific choice no longer matters.
The Fock modes then obey boson
commutation relations
\begin {equation}
     \bigl[ b_n , b^{\dag}_m \bigr] =
     \bigl[ d_n , d^{\dag}_m \bigr] =
     \delta_{n}^{m} \ , \quad\  {\rm and} \quad
     \bigl[ b_n , d_m \bigr] =
     \bigl[ b_n , d^{\dag}_m \bigr] = 0
\;. \label{focomrel} \end    {equation}
Finally one notes that a large gauge transformation
$z\rightarrow z + 1$ produces only $m_0 \rightarrow m_0 + 1$
and thus only a change of the overall phase in Eq.(\ref{fockexp}).
Most importantly it does {\it not} change the particle-\-hole
assignment and thus the Fock vacuum defined with respect to
$b_m$ and $d_m$ is {\it invariant} under these transformations.

The only modes not discussed thus far are the $A^-$ fields.
As is known for the Coulomb gauge, they are redundant variables
obtained from implementing Gauss' law strongly.
Explicitly, Gauss' law reads
\begin {equation}
   -\partial_-^2  A_3 =  g  J^+_3, \qquad\quad
   -(\partial_- + i g   A^+_3)^2  A_+
   = g  J^+_+
\ , \label {gausscomp} \end   {equation}
and the hermitian conjugate of the latter with
$(J^+_+)^{\dag} \equiv J^+_-$.
The currents are, explicitly,
\begin {equation}
  J^+_3 = {1\over i} \bigl(
           \varphi_+ \pi_- - \varphi_- \pi_+ \bigr)_s
 \ \  {\rm and}\quad
  J^+_+ = {1\over i} \bigl(
           \varphi_3 \pi_+ - \varphi_+ \pi_3 \bigr)_s
\ . \end    {equation}
The index $s$ indicates noncommuting operators in this product
which must be symmetrised in order to preserve hermiticity.
The Eqs.(\ref{gausscomp}) are trivially soluble for the
fields $A^-$ subject to the following exception.
The first of Eq.(\ref{gausscomp})  can be solved
only if the zero mode $ \langle J^{+}_{3} \rangle_\circ \equiv  Q_0 $
on the r.h.s vanishes.
This cannot be satisfied as an operator, but must be used to
select out physical states, {\it i.e.}\
$     Q_0 \vert {\rm {phys}}  \rangle \ \equiv 0 $.
In second-quantised form this gives
\begin {equation}
      Q_0 \vert {\rm {phys}} \rangle =
   \sum_{m = {1\over 2}}^{\infty}
   \Bigl(b^{\dag}_m   b_m - d^{\dag}_m   d_m \Bigr)
    \vert {\rm {phys}} \rangle = 0 \ .
\label{gaussop}
\end    {equation}
States satisfying this
have the same total number of ``b'' and ``d'' particles.
The resemblance to the electric-charge neutrality condition
is because the residual global gauge symmetry group factored out
of the Hilbert space is Abelian.
Evidently, the b-modes carry ``charge'' 1, the d-modes
charge -1 and the a-modes charge 0.

\section{A Diagrammar for the Zero Mode Problem}
\begin{figure}
\vspace{1em}
\centering
\begin{tabular}{|c|c|}\hline
{\bf Graph} & {\bf Diagram Rules} \\ \hline\hline
\parbox{1in}{\vspace{0.2cm}\psfig{figure=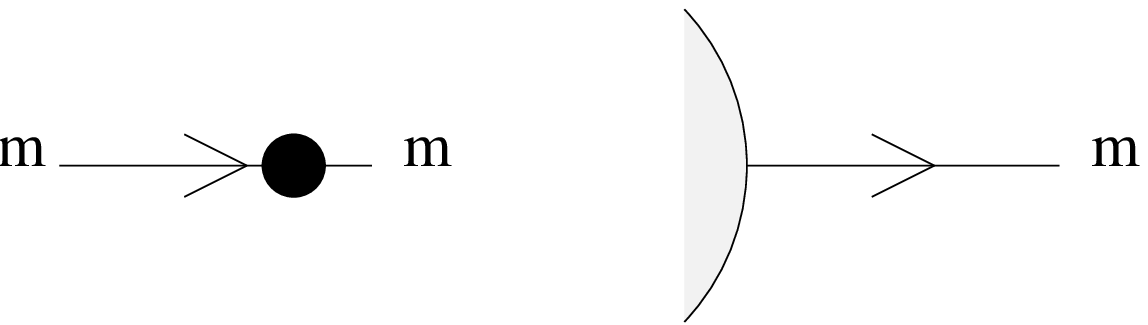,height=1.0cm} \vspace{0.2cm}
}
& { $ \Delta_m(\zeta) \; ,  \;
b^{\dag}_m |0\rangle $ }
\\ \hline
\parbox{1in}{\vspace{0.2cm}\psfig{figure=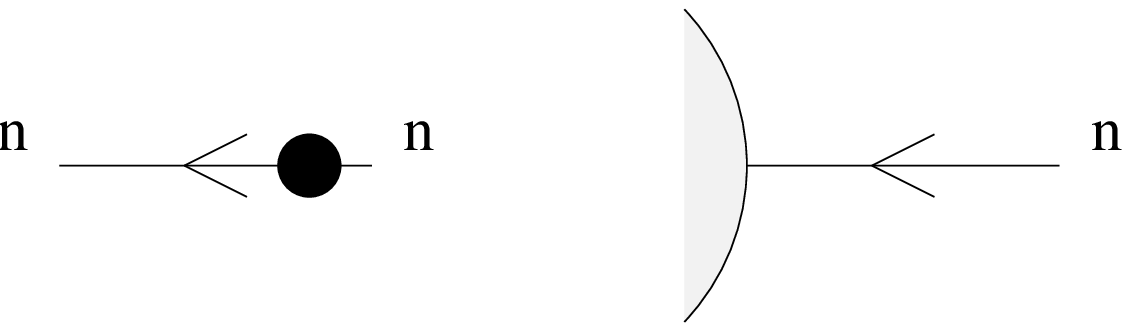,height=1.0cm}
\vspace{0.2cm}}
& $ \Delta_n(-\zeta) \; ,\;
d^{\dag}_n |0\rangle $
\\ \hline
\parbox{1in}{\vspace{0.2cm}\psfig{figure=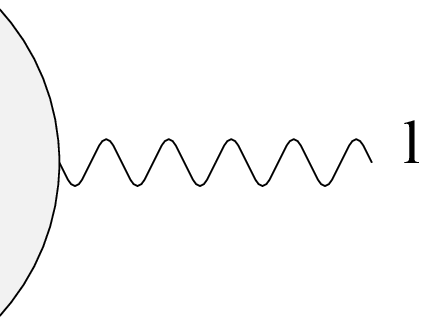,height=1.0cm}\vspace{0.2cm}}
&  $ a^{\dag}_l |0\rangle $
\\ \hline
\parbox{1in}{\vspace{0.2cm}\psfig{figure=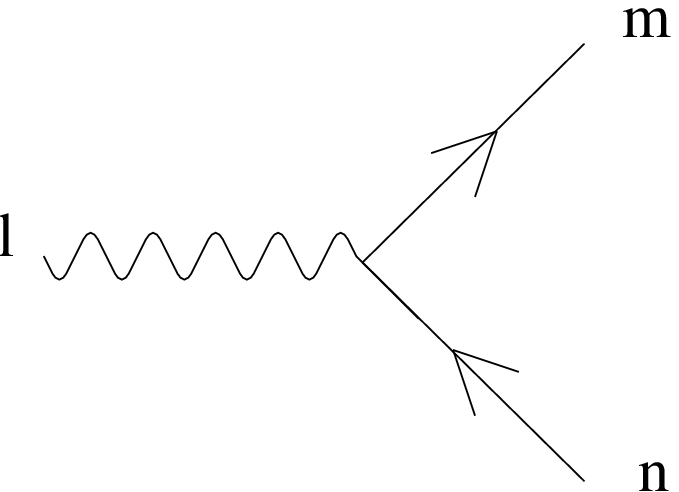,height=2cm} \vspace{0.2cm}}
&  $ \Gamma_0^{lmn}(\zeta) $
\\ \hline
\parbox{1in}{\vspace{0.2cm}\psfig{figure=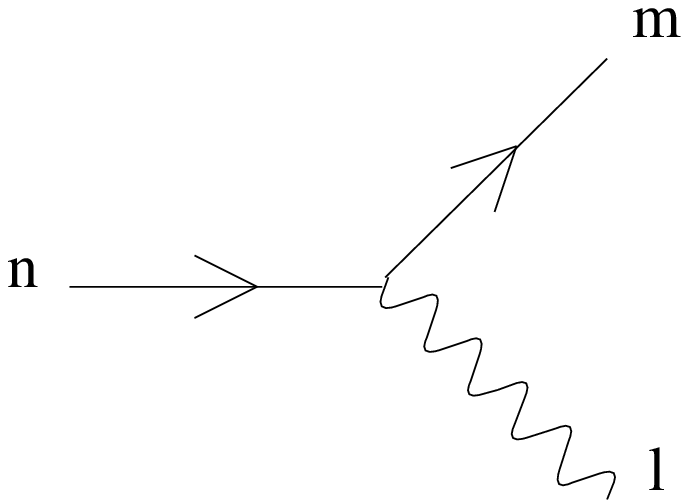,height=2cm} \vspace{0.2cm}}
&  $ \Gamma_1^{lmn}(\zeta) $
\\ \hline
\parbox{1in}{\vspace{0.2cm}\psfig{figure=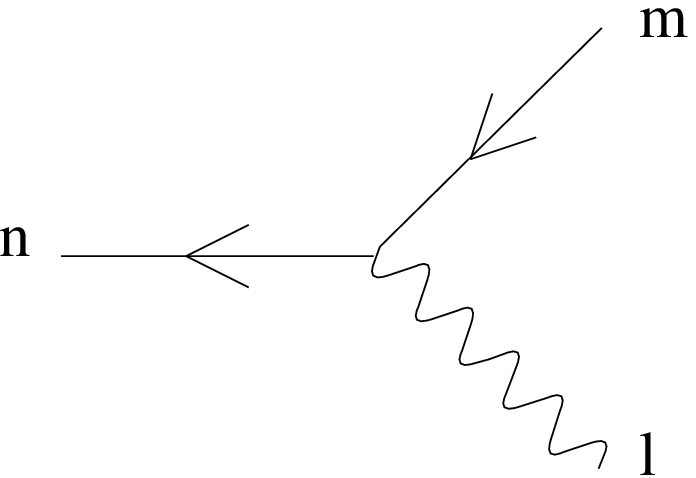,height=2cm} \vspace{0.2cm}}
& $ \Gamma_1^{lmn}(-\zeta) $
\\ \hline
\parbox{1in}{\vspace{0.2cm}\psfig{figure=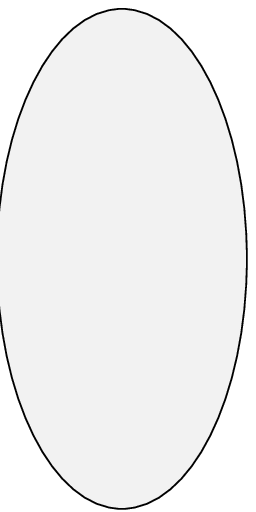,height=1cm}\vspace{0.2cm}}
&  $ a_0 $
\\ \hline
\end{tabular}
\vspace{1em}
\caption{Diagrammatic Rules for the objects appearing in the
constraint equation Eq.(\protect\ref{theconstraint}). Hermitian conjugation
involves reflection through the vertical plane without
changing senses of arrows. Further rules
are explained in the text.}
\label{fig:rules}
\end{figure}

The origin of the constraint equation for $a_0$
has already been discussed in \cite{PKP95}:
the zero mode of the diagonal part of
Eq.(\ref{eqofmot}),
$  8\pi {\rm Tr} \ \langle \tau ^3
({\bf D}^\alpha {\bf D}_\alpha + \mu_0^2)\Phi \rangle_\circ /g^2 =0.$
After some algebra this gives
\begin {equation}
   i  \Bigl\langle \
       \varphi_+ {1\over (\partial_- - i  g   v)} J^+_-
     - \varphi_- {1\over (\partial_- + i  g   v)} J^+_+
     \ \Bigr\rangle_{0,s}  - {{\mu_0^2}\over {g^2 \sqrt{4\pi} }} a_0 = 0 \ .
\end    {equation}
In the absence of a mass term, the constraint in its
`full glory' was shown in \cite{Kal94}.
To proceed here I first introduce the dimensionless ratio
of boson mass and coupling
\begin{equation}
\rho_0 \equiv {{4\pi \mu_0^2}\over{g^2}}
\;.
\end{equation}
Next I introduce the
following notation:
\begin{eqnarray}
\Delta_m(\zeta) & = & {1\over {2 (m + \zeta)}}
\nonumber \\
\Gamma_0^{lmn} (\zeta) & = &
[( {{w_l}\over{v_n}} + {{v_n}\over{w_l}} ) u^3_m
+
( {{w_l}\over{u_m}} + {{u_m}\over{w_l}} ) v^3_n ]
\nonumber \\
\Gamma_1^{lmn} (\zeta) & = &
-
[( {{w_l}\over{u_m}} - {{u_m}\over{w_l}} ) u^3_n
+
( {{w_l}\over{u_n}} + {{u_n}\over{w_l}} ) u^3_m ]
\; .
\end{eqnarray}
The constraint equation turns out to be
\begin{eqnarray}
      \sum_{n={1\over2}}^{\infty}[ \Delta_n(\zeta)  (b^{\dag}_n b_n a_0)_s  +
      \Delta_n(-\zeta) (d^{\dag}_n d_n a_0)_s ]
 + \rho_0 a_0  & = & \nonumber \\
  \sum_{l=1}^\infty \sum_{m,n={1\over2}}^{\infty}
 [
\delta^l_{m+n}
\Gamma_0^{lmn}(\zeta)
(a^{\dag}_l b_m d_n  +  a_l b^{\dag}_m d^{\dag}_n)
+
 \nonumber \\
\delta^n_{l+m}
\Gamma_1^{lmn}(\zeta)
(a^{\dag}_l b^{\dag}_m b_n \ + \ a_l b_m b^{\dag}_n)
\;  +  \;
\delta^n_{l+m}
\Gamma_1^{lmn}(-\zeta)
(a^{\dag}_l d^{\dag}_m d_n & + &  a_l d_m d^{\dag}_n)
] .
\label{theconstraint}
\end{eqnarray}
One observes that, the dynamically independent composite
operators on the left-hand side of the equation
are two-body operators -- propagators --
while on the right-hand side they are three-body
operators, namely interaction vertices.
This suggests diagrammatic rules for the various
quantities in this expression which are represented in
Fig.(\ref{fig:rules}).
Additional rules are as follows:
\begin{enumerate}
\item Matrix elements are composed by attaching legs to the
right and left of the blob representing the operator.
\item Closed loops represent a sum over all positive
integer or half-integer momenta.
\item Detached diagrams represent multiplication of the corresponding
expressions.
\item Hermitian conjugation is achieved by reflection of the diagram
while keeping the sense of arrows the same.
\item Charge conjugation is achieved by reversing the direction
of the arrows.
\end{enumerate}
Evidently, these are quite similar to Feynman rules but
describe the building of more than just S-matrix elements.
Here I will use them to relate matrix elements of the
constrained zero mode as determined by the constraint equation.

\begin{figure}
\vspace{1em}
\centerline{
\parbox{1in}{
\vspace*{0.2cm}\psfig{figure=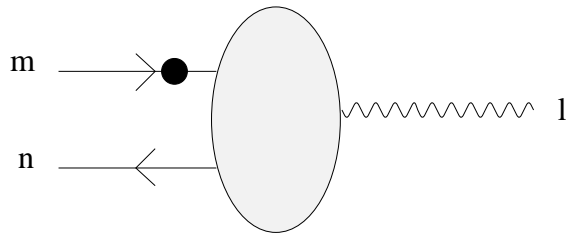,height=3cm}
\vspace*{0.2cm} }}
\vspace{1em}
\caption{Example of use of the diagrammatic rules
for the product of a matrix element and a propagator:
$\Delta_m(\zeta) \langle 0 | b_m d_n \ a_0 \ a^{\dag}_l |0 \rangle$.}
\label{fig:matrixel}
\end{figure}

To illustrate the use of the rules, the object
\begin{equation}
\Delta_m(\zeta) \langle 0 | b_m d_n \ a_0 \ a^{\dag}_l |0 \rangle \ ,
\end{equation}
describes  a matrix element of $a_0$ multiplied
by a propagator factor.
Diagrammatically, this is represented by Fig.(\ref{fig:matrixel}).

\begin{figure}
\vspace{1em}
\centerline{
\protect\parbox{1in}{
\vspace*{0.2cm}\psfig{figure=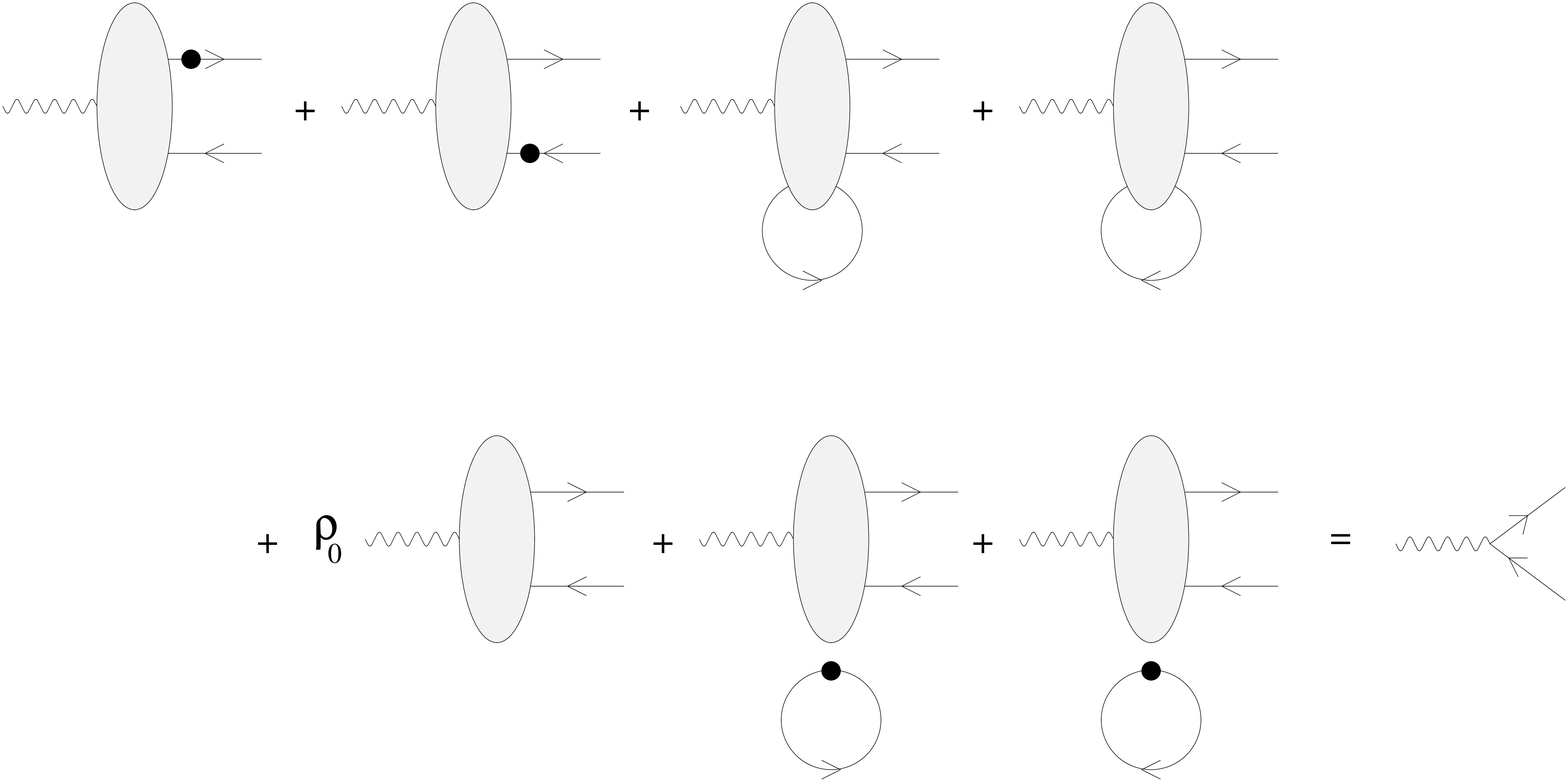,width=13cm}
\vspace*{0.2cm}} }
\vspace{1em}
\caption{The constraint equation Eq.(\protect\ref{theconstraint}) evaluated
between $\langle 0| a_l$ and $b^{\dag}_m d^{\dag}_n |0\rangle$ depicted
diagrammatically.}
\label{fig:constreqn}
\end{figure}

Next, one can represent matrix elements of the
constraint equation in a particular sector. For example,
taking $\langle 0| a_l $ on the left and $b^{\dag}_m d^{\dag}_n |0\rangle$
on the right and using commutators where possible,
one obtains an equation which can be represented
diagrammatically as shown in
Fig.(\ref{fig:constreqn}).
The mass term is evidently the fifth term.
Similar diagrammatic equations can be generated by sandwiching
the constraint between higher particle states, and the intuitive
principal for their construction is straightforward:
\begin{enumerate}
\item On the right hand
side, join up the incoming and outgoing lines with the three available
vertices in all possible permutations, with the remaining lines
running through as `spectators'.
\item On the left hand side, assign
propagator legs once only to all `incoming/outgoing' b- or d-boson legs,
attach all possible permutations of tadpoles and loops.
\end{enumerate}

The following features can be observed in Fig.(\ref{fig:constreqn}) and
the original equation Eq.(\ref{theconstraint}).

Firstly, the equation involves matrix elements of the
constrained {\it mode} in the 5-body sector (two
particles `in', three `out'). To get at these one
takes matrix elements of the constraint {\it equation}
in this higher sector. This I will show below.
Naturally, this in turn involves matrix elements from the 7-body
sector (three `in', four `out').
This sequence carries on {\it ad infinitum}.
Thus all the particle sectors appear related to each other
via the constrained mode which is playing the role of an effective
interaction. The question is how complicated is this coupling of
sectors?

One can rule out any coupling of the vacuum
sector into the hierarchy. The VEV $\langle 0 | a_0 | 0 \rangle = 0$
because under charge conjugation\footnote{For a colour matrix ${\bf{\Phi}}$,
charge conjugation is represented by $\Phi_{ij} \rightarrow
- \Phi{ji}$, where $(i,j)$ are the indices of the matrix
in colour space \cite{Dal95}. } $a_0 \rightarrow - a_0$
while the vacuum is even.

In addition, the following commutation relations for $a_0$ must be
satisfied:
\begin{eqnarray}
  \left[ a_0, Q_0 \right] & = & 0 \nonumber \\
  \left[ a_0, P^+ \right] & = & 0 \nonumber .
\end{eqnarray}
These can be used to further reduce the number of non-zero matrix elements.

\begin{figure}
\vspace{1em}
\centerline{
\parbox{1in}{
\vspace*{0.2cm}\psfig{figure=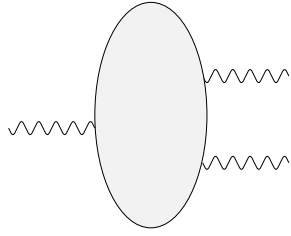,height=3cm}
\vspace*{0.2cm} }}
\vspace{1em}
\caption{An example of a matrix element vanishing because the
operator cannot be replaced by one of the basic vertices in
the constraint equation.}
\label{fig:vanishing}
\end{figure}

Momentum conservation through the vertices will
mean that for many sectors the right-hand side of the
hierarchy of coupled equations will vanish. It is difficult
to prove generally, but the system of linear coupled equations
can be shown to be nonsingular for simple examples. Thus
matrix elements of $a_0$ which cannot be described in terms
of the basic vertices can be argued to vanish. An example
is the matrix element in Fig.(\ref{fig:vanishing}).

\underline {Renormalisation of the Constraint.}
The graphs in Fig.(\ref{fig:constreqn}) with detached bubbles are
logarithmically divergent.
I introduce a cutoff regulator $\Lambda$ in the momentum sums
so that these terms become
\begin{equation}
 \sum_{n=1/2}^{\Lambda} ( \Delta_n(\zeta)  +
       \Delta_n(-\zeta) ) a_0  \sim (\ln \Lambda  + {\rm{finite}} ) a_0
\;.
\end{equation}
This divergence can be absorbed into the mass term
by a subtraction of the sum for $\zeta=0$, namely with
physical mass $\rho$ given by
\begin{eqnarray}
\rho & = & \rho_0 + 4 \sum_{m={1\over2}}^{\Lambda} \Delta_m(0) \nonumber \\
& = & \rho_0 + 4[ {1\over2} (\gamma_{E} + \ln\Lambda) + \ln 2
            + {\cal O}(1/\Lambda) ]
\;.
\label{renmass}
\end{eqnarray}
One can verify that this is the same renormalisation used to
remove all cutoff dependence in the two-particle bound state equation
in, for example, \cite{PaB95}.
A similar relationship between
the two renormalisations has already been observed in the
two-dimensional $\phi^4$ theory by \cite{BPV93}.
As a consequence, the left hand side of the original constraint
now takes the following form
\begin{equation}
\sum_{m=\frac{1}{2}}^{\Lambda}
\bigl[ \Delta_m(\zeta) \bigl(
b^{\dag}_m b_m a_0 + b^{\dag}_m a_0 b_m + a_0 b^{\dag}_m b_m
+ (b_m a_0 b^{\dag}_m - a_0 ) \bigr)
+ (b \rightarrow d, \zeta \rightarrow - \zeta) \bigr]
+ {\hat{\rho}} (\zeta) a_0
\end{equation}
with
\begin{equation}
\hat{\rho}(\zeta) = \rho + 2 \sum_{m={1\over2}}^\Lambda
        (\Delta_m(\zeta) + \Delta_m(-\zeta) - 2 \Delta_m(0) )
\label{renormal}
\end{equation}
being a convergent function in $\Lambda$.
In this way, matrix elements of the zero mode $a_0$ are
rendered cutoff independent.

\begin{figure}
\vspace{1em}
\centerline{
\parbox{1in}{
\vspace*{0.2cm}\psfig{figure=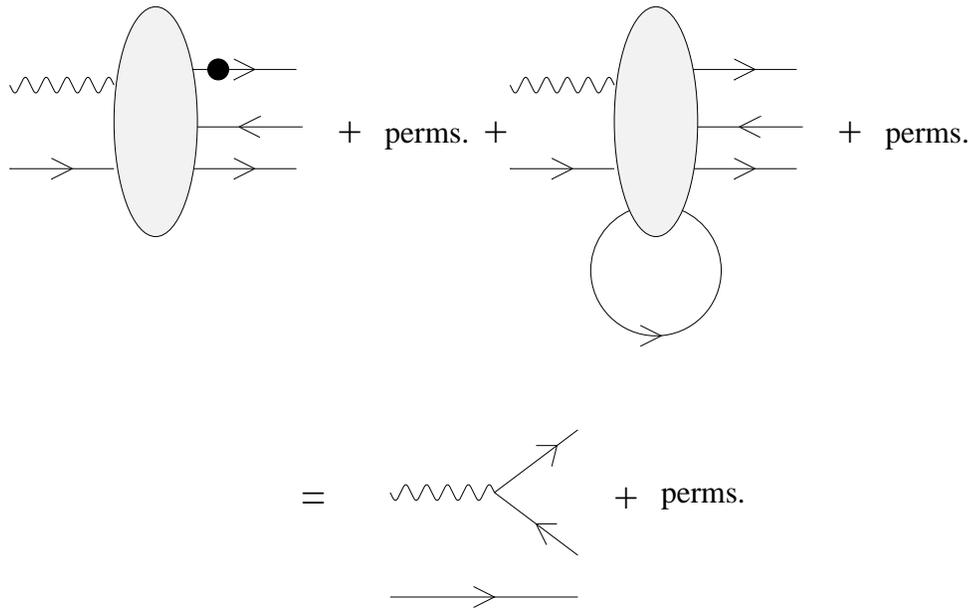,height=8cm}
\vspace*{0.2cm} }}
\vspace{1em}
\caption{Some of the terms in one of the 5-particle sector of
the constraint equation for states
$\langle 0| a_l b_m$ and $b^{\dag}_{m_1} d^{\dag}_{n} b^{\dag}_{m_2} | 0
\rangle$.}
\label{fig:5particle}
\end{figure}

\underline{5-Particle Sector.}
I now examine what happens in a higher particle sector.
The equation is represented schematically in Fig.(\ref{fig:5particle})
for the 5 particle
sector when evaluating the constraint between states
$\langle 0| a_l b_m$ and $b^{\dag}_{m_1} d^{\dag}_{n} b^{\dag}_{m_2} | 0
\rangle$.
Evidently, the right hand side of the equation is non-zero only
for particular configurations of momenta in the left hand side.
Namely, for the case when the zero mode is {\it not} acting as
essentially just a three-point operator with some spectators,
the right hand side is zero.
Thus, basically the constrained zero mode is
a three-point vertex. However, because of terms such as the third and fourth
in Fig.(\ref{fig:constreqn}) it is not just the
bare vertex. Rather the vertices are dressed.

\begin{figure}
\vspace{1em}
\centerline{
\parbox{1in}{
\vspace*{0.2cm}\psfig{figure=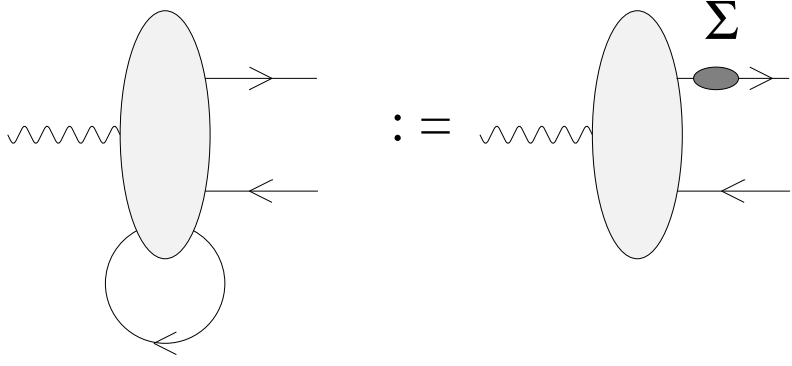,height=6cm}
\vspace*{0.2cm} }}
\vspace{1em}
\caption{Identification of terms that connect different particle
sectors of the constraint with an effective operator with
insertion $\Sigma(\zeta)$.
The analogous diagram can be drawn for the anticlockwise flow of
momentum in the loop on the left hand side
but where the insertion is made on
the lower $d$-leg. Namely $\Sigma$ appears with argument $-\zeta$. }
\label{fig:insertion}
\end{figure}

\underline{Iteration to all orders.}
One can attempt solving the equation iteratively --
essentially a weak coupling expansion.
One would observe that to any order in the
iteration the three separate contributions coming from
respectively $\Gamma_0(\zeta)$ and $\Gamma_1(\pm\zeta)$
always decouple from each other. Mixing occurs solely via
contractions between respectively $b$ with $b^{\dag}$ and
$d$ with $d^{\dag}$.
Diagrammatically, one can picture this in terms of precisely the terms
in Fig.(\ref{fig:constreqn}) that make the
constraint operatorially non-trivial,
namely the third and fourth terms which
connect the different particle sectors.
The way I proceed is to reexpress such terms as an effective
{\it three-body} operator with an {\it insertion}, $\Sigma$, on the leg
of the same colour as that in the loop. This is illustrated
in Fig.(\ref{fig:insertion}).
Evidently, no mixing can occur between $b$ and $d$ modes
once the basic three-point vertices have been extracted.
Thus the insertion can only arise as the
sum to all orders
of products of the basic propagator for a given value of $\zeta$
as follows:
\begin{equation}
\Sigma_m(\zeta) = \lim_{N\rightarrow\infty} \sum_{r=1}^N ( \Delta_m(\zeta) )^r
=
\lim_{N\rightarrow\infty}
{{1 - \Delta_m(\zeta)^N} \over {1 - \Delta_m(\zeta)}}
\;.
\label{sumup}
\end{equation}
Observe that for all but the lowest mode $m=1/2$, the propagator
satisfies $|\Delta_m(\zeta)| < 1$. Thus for the higher modes,
Eq.(\ref{sumup}) can be further simplified. However, for momentum
$m=1/2$ the propagator will in general be `large', and thus in turn
the effect of $\Sigma_{1/2}(\zeta)$ will be even larger.

\underline{Solution.}
The above considerations motivate the following operator ansatz for
the solution
\begin{eqnarray}
a_0 & = & \sum_{l,m,n} [
 C_0^{lmn}(\zeta) \delta^l_{m+n} ( a^{\dag}_l b_m d_n + \; {\rm{h.c.}} )
\nonumber \\
& + &  C_1^{lmn}(\zeta) \delta^n_{m+l} ( b^{\dag}_n b_m a_l + \; {\rm{h.c.}} )
+ C_1^{lmn}(- \zeta)  \delta^n_{m+l} ( b \rightarrow d) ]
\; .
\label{thesolution}
\end{eqnarray}
This expression by construction satisfies the symmetries
obeyed by the constrained zero mode.
It is essentially a three-body operator, where the dressing of the
vertices will be absorbed into the, as yet, arbitrary coefficients
$C_a^{lmn}(\zeta)$.
I insert Eq.(\ref{thesolution}) into Eq.(\ref{theconstraint})
with the identification of Fig.(\ref{fig:insertion}) implemented.
It suffices to consider the constraint in the three-particle
space in order to solve for the coefficients, once one builds in
the iteration to all orders in the insertion via Eq.(\ref{sumup}).
I represent the constraint as $L[a_0] = R$,
where $L$ and $R$ denote the left- and right-hand sides of
Eq.(\ref{theconstraint}). I take the matrix elements
\begin{eqnarray}
\langle 0 | a_l \ L[a_0] \ b^{\dag}_m d^{\dag}_n |0 \rangle & = &
\langle 0 | a_l \ R \ b^{\dag}_m d^{\dag}_n |0 \rangle \;,
\nonumber \\
\langle 0 | b_n \ L[a_0] \  b^{\dag}_m a^{\dag}_l |0 \rangle & = &
\langle 0 | b_n \ R \ b^{\dag}_m a^{\dag}_l |0 \rangle \;.
\end{eqnarray}
The d-mode matrix element gives no additional information because of
charge conjugation symmetry.
The two equations are decoupled and one can
straightforwardly determine the $C_0$ and $C_1$ coefficients in terms
of the fundamental vertices and propagator
\begin{eqnarray}
C_0^{lmn}(\zeta) & = &
{ {\Gamma_0^{lmn}(\zeta)} \over { {\cal D}^{(0)}_{m,n}(\zeta) } },\;
{\cal D}^{(0)}_{m,n}(\zeta) \equiv
\Delta_m(\zeta) + \Delta_n(-\zeta) + \Sigma_m(\zeta) + \Sigma_n(-\zeta)
+ \hat{\rho}(\zeta),
\nonumber \\
C_1^{lmn}(\zeta) & = &
{ {\Gamma_1^{lmn}(\zeta)} \over  { {\cal D}^{(1)}_{m,n}(\zeta) } }, \;
{\cal D}^{(1)}_{m,n}(\zeta) \equiv
\Delta_m(\zeta) + \Delta_n(\zeta) + \Sigma_m(\zeta) + \Sigma_n(\zeta)
+ \hat{\rho}(\zeta)
\;.
\label{coeffs}
\end{eqnarray}
With the coefficient functions determined one can now
directly give the lowest particle matrix elements
as
\begin{eqnarray}
\langle 0|a_L \ a_0 \ b_M^{\dag} d_N^{\dag} |0\rangle & = &
\delta^L_{M+N} C_0^{LMN}(\zeta) \nonumber \\
\langle 0|  b_{M_1} \ a_0 \ b^{\dag}_{M_2} a^{\dag}_L |0\rangle & = &
\delta^{M_1}_{{M_2}+L} C_1^{L M_2 M_1}(\zeta) \nonumber \\
\langle 0|  d_{N_1} \  a_0 \  d^{\dag}_{N_2} a^{\dag}_L |0\rangle & = &
\delta^{N_1}_{{N_2}+L} C_1^{L N_2 N_1}(-\zeta)
\;.
\label{matrelements}
\end{eqnarray}
Eqs.(\ref{thesolution}), (\ref{coeffs}) and (\ref{matrelements}) are the
main results of this section.

\section{Revisiting the `Adiabatic Potential' }
The Hamiltonian is the Poincar{\'e} generator $P^-$,
given in Appendix D with other details from \cite{PKP95}.
After removing a dimensionful
factor $L(g/{4\pi})^2$ one obtains the rescaled Hamiltonian, $H$.
One seeks the ground state of the theory.
If not for the gauge mode $\zeta$, the naive arguments leading to
the trivial Fock vacuum being the true ground state would
hold rigorously. Now in fact the ground state is some
infinite superposition of $\zeta$ zero modes. One could try
representing this in an oscillator basis but the true ground
state is mostly likely some coherent state of these oscillator
modes.
As mentioned, I use instead a Schr\"odinger representation
within the adiabatic approach of \cite{PKP95,LST94}.
The gauge mode $\zeta$ is frozen for the purposes of computing
the ground state in the `particle sector'. It is here the advantages
of the light-cone become significant: the ground state is now the
Fock vacuum. However, the tadpole terms that arise as one normal orders
the Hamiltonian with respect to the trivial vacuum generate
$\zeta$-dependent structures which become quantum operators once
the gauge mode is unfrozen,
\begin{equation}
H = \; :\!H[\zeta]\!: + V[\zeta]
\;.
\end{equation}
Equivalently, $V[\zeta] \equiv \langle 0| H[\zeta] |0\rangle$.
As shown in \cite{PKP95} and Appendix D,
there is also a kinetic term for $\zeta$ surviving in $:\!H\!:$.
Thus, projecting in the Fock vacuum sector,
we obtain a Hamiltonian
\begin{equation}
H_\zeta = - 4 {{d^2}\over{d\zeta^2}} + V[\zeta] \;.
\label{schroed}
\end{equation}
For later purposes it is convenient to break $V$
into three terms, coming respectively from the mass term, the
constrained mode $a_0$ {\it de}pendent term and
the $a_0$ {\it in}dependent term
\begin{equation}
V[\zeta] = V_{\rho}[\zeta] + V_a[\zeta] + V_0[\zeta]
\;.
\end{equation}
Solving the Schr\"odinger equation corresponding to $H_\zeta$
generates the ground state wave functional
$\Psi_0(\zeta)$ allowing the introduction of the `true' ground state
as the tensor product state
\begin{equation}
|\Omega\rangle \equiv \Psi_0[\zeta] \otimes |0\rangle
\;.
\end{equation}

\underline {Vacuum `Degeneracy'.}
The only residue of the large gauge symmetries, Weyl reflections
and central conjugations in the fundamental modular domain
is the symmetry $\zeta \rightarrow - \zeta$, which is the basic symmetry
of the potential. Thus there is an associated quantum number
which labels the wavefunctions, namely symmetric $\Psi_n^{(+)}[\zeta]$
and antisymmetric $\Psi_n^{(-)}[\zeta]$ states.
Thus identical copies of the particle
spectrum can be built on either the vacuum described by
$\Psi_0^{(+)}[\zeta]$ or $\Psi_0^{(-)}[\zeta]$ differing only by
a fixed shift in the energy. The shift will be finite as the
longitudinal interval length $L$ (factored out of $H_\zeta$
at this point) is kept finite.
This is how the $Z_2$ analogue of $\theta$-vacua arise in this theory.
The finiteness of the shift as $L \rightarrow \infty$ depends on the
specific structure of $V$. In particular, the presence of an
impenetrable barrier in the {\it centre} of the fundamental
domain would bring the two lowest energies into exact degeneracy
for any interval size $L$.

\underline {Renormalisation of $V_{\rho} + V_0$.}
The term $V_0$ was computed in \cite{PKP95}.
Here I quote the relevant result:
\begin{equation}
 V_0 [\zeta]  = \sum_{n ={1\over2}}^\Lambda
           \sum_{m ={1\over2}}^\Lambda
  \Bigl({u_m \over  v_n}
         - {v_n \over  u_m} \Bigr)^2  w^4_{m + n}
         + \sum_{l = 1}^{\Lambda + \frac{1}{2} }
         \sum_{m ={1\over2}}^\Lambda
  \Bigl[ \Bigl({w_l \over  v_m}
         - {v_m \over  w_l} \Bigr)^2  v^4_{m + l}
  + \Bigl({w_l \over  u_m}
         - {u_m \over  w_l} \Bigr)^2  u^4_{m + l} \Bigr]
\end{equation}
where $\Lambda$ is the (half-integer valued) cutoff in ultra-violet
momenta.
Using $ u_n(\zeta)=1/\sqrt{n+\zeta}$, $ v_n(\zeta)=1/\sqrt{n-\zeta}$, and
$ w_n = 1/\sqrt{n}$ one can check this has the symmetry
$\zeta \rightarrow - \zeta$. Substituting these coefficients, the
expression is
\begin{eqnarray}
 V_0 [\zeta] & = &      \sum_{n ={1\over2}}^\Lambda
           \sum_{m ={1\over2}}^\Lambda
 {{((m + \zeta) - (n - \zeta))^2} \over {(m + n)^2 (m + \zeta) (n - \zeta)}}
        \nonumber \\
             & + & \sum_{l = 1}^{\Lambda + \frac{1}{2} }
           \sum_{m ={1\over2}}^\Lambda
\Bigl( { {(m - l + \zeta)^2} \over {l (m + \zeta) (l + m + \zeta)^2}}
              +
{ {(m - l - \zeta)^2} \over {l (m - \zeta) (l + m - \zeta)^2}}
                                                       \Bigr) \;.
\label{v0sums}
\end{eqnarray}
Closer inspection reveals the sums to be logarithmic in the
cutoff with a coefficient that depends on $\zeta$. It is this
dependence I seek to extract.
A useful trick is to add and subtract each of the three terms in
Eq.(\ref{v0sums}) evaluated at $\zeta=0$. It is then a straightforward
though tedious task to show that the expression can be
reorganised as
\begin{equation}
 V_0 [\zeta] = 4 G[\zeta] \sum_{m=\frac{1}{2}}^\Lambda {1\over m}
 + {\rm {Convergent}}[\zeta] + {\rm{const}}
\label{v0div}
\end{equation}
with
\begin{equation}
   G[\zeta] =  \sum_{m=\frac{1}{2}}^{\Lambda}
({1\over{m+\zeta}} + {1\over{m-\zeta}} - {2\over m} )
\;.
\end{equation}
One recognises in $G$ the convergent functional $\hat{\rho}(\zeta) - \rho$,
Eq.(\ref{renormal}).
One sees it comes in as the operator dependence of the logarithmic divergence.

The contributions of the mass term are more straightforward.
After normal ordering, one extracts the structure
\begin{equation}
V_{\rho}[\zeta] = \rho_0 (
           \sum_{l=1}^\Lambda {1 \over l}
         +  \sum_{m=\frac{1}{2}}^\Lambda {1 \over {(m + \zeta)}}
         +  \sum_{m=\frac{1}{2}}^\Lambda {1 \over {(m - \zeta)}}
                         )
\label{masspot}
\end{equation}
where $\rho_0$ was the bare (dimensionless) mass.
In order to render the constraint equation and particle sector
convergent and cutoff independent, the renormalised mass
$\rho = \rho_0 + 2 \sum_{m=1/2}^\Lambda {1\over m}$ was introduced,
Eq.(\ref{renmass}).
Substituting this into Eq.(\ref{masspot}) gives
\begin{equation}
V_{\rho}[\zeta] = - 4 G[\zeta] \sum_{m=\frac{1}{2}}^\Lambda {1\over m}
 + \rho G[\zeta] + {\rm{const.}}
\;.
\label{rhodiv}
\end{equation}
The $\zeta$-dependent but logarithmic divergent terms in Eq.(\ref{v0div})
and Eq.(\ref{rhodiv}) are equal
but opposite in sign. Thus the same renormalisation holds
respectively in the three separate cases of the two-particle
bound state equation, the constraint equation, and in the
$V_0 + V_{\rho}$ parts of the potential.

As shown in \cite{PKP95}, the shape of this part of the potential
is flat within the fundamental domain, rising to positive
infinity at the domain boundaries $\zeta = \pm \frac{1}{2}$.
A careful study of the nature of the singular behaviour
as $\zeta \rightarrow \pm \frac{1}{2}$
suggests it is not as strong as $1/(\zeta \pm \frac{1}{2})^2$,
thus the potential at the domain boundaries
appears to be penetrable. However,
as argued in \cite{Lus83,LST94,LNT94}, the wavefunctions should be supplemented
with a Jacobian determinant coming from the Faddeev-Popov structure
in the present gauge. The wavefunctions are thus
forced to vanish rigorously at the
boundaries of the fundamental modular region.
Thus the spectrum is approximately
that of a square well which itself corresponds to that of pure glue SU(2)
theory on a cylinder \cite{Het93},
with a sinusoidal ground state wavefunction.
Either way, in the absence of the constrained mode there is no
structure within the centre of the fundamental modular region.

\section{Impact of the Constrained Mode}
I now include the effects of $a_0$.
As shown in \cite{PKP95},
there are terms in the Hamiltonian linear and quadratic in $a_0$,
which were separated into $ H_{Constr} = H_{Constr}^{(1)} +  H_{Constr}^{(2)}
$.
They were respectively
\begin {eqnarray}
   H_{Constr}^{(1)} & = &
   2 \sum_{k ={1\over2}}^{\infty} \Bigl[ %
      u_k^3\ \Bigl(B^{\dag}_k   Q_- (k) +
                      Q_-^{\dag}(k) B_k \Bigr)_s
  + v_k^3\ \Bigl(D^{\dag}_k   Q_+ (k) +
                      Q_+^{\dag}(k) D_k \Bigr)_s \Bigr]
, \nonumber  \\
    H_{Constr}^{(2)} & = & \sum_{k ={1\over2}}^{\infty} \Bigl[\ %
      u_k^2\ \bigl(B_k B_k^{\dag} + B_k^{\dag} B_k \bigr)
  + v_k^2\ \bigl(D_k D_k^{\dag} + D_k^{\dag} D_k \bigr)
\ \Bigr] \ , \end   {eqnarray}
%
where the $Q_\pm$ arise from the $a_0$ independent
parts of the currents and are given in Appendix D.
The other operators are just $B_k = (a_0 b_k + b_k a_0)/2$
and $D_k = (a_0 d_k + d_k a_0)/2$ which arise because
of symmetrisation of noncommuting operator products.
I correspondingly consider the VEVs of the two terms separately.
The linear term leads to
\begin{eqnarray}
V_1[\zeta,\rho] & \equiv & \langle 0|  H_{Constr}^{(1)} |0\rangle
   \nonumber \\
                & = &  {1\over2} \sum_{l,m,n} \delta^n_{m+l} [
R^{lmn}(\zeta) ( \langle 0| a_l b_m a_0 b^{\dag}_n |0 \rangle
            + \langle 0| b_n a_0 a_l^{\dag} b_m^{\dag} |0 \rangle )
\nonumber \\
& + & R^{lmn}(-\zeta) ( b \rightarrow d )]
\label{linearvaccontr}
\end{eqnarray}
with
\begin{equation}
R^{lmn}(\zeta) = u_n^3 ({{w_l} \over {u_m}} - {{u_m} \over {w_l}} )
\label{partvertex}
\end{equation}
which is actually part of the vertex function $\Gamma_1(\zeta)$.
The quadratic term gives
\begin{eqnarray}
V_2[\zeta,\rho] & \equiv & \langle 0| H_{Constr}^{(2)} |0\rangle
\nonumber \\
                  & = & \sum_{m}  [
\Delta_m(\zeta) \langle 0| b_m a_0^2 b^{\dag}_m |0 \rangle
     +
\Delta_m(-\zeta) \langle 0| d_m a_0^2 d^{\dag}_m |0 \rangle ]
\;.
\end{eqnarray}

One sees that the relevant contributions to the VEV come purely from the lowest
particle sector matrix elements of $a_0$. For $V_2$ this follows after
inserting a complete set of states between the two $a_0$ operators.

It is next a tedious task of inserting the solutions as given above.
The most compact expression for the result is
\begin{eqnarray}
\langle 0 | H_{Constr} | 0 \rangle =  & {} &
- 2 \sum_{l=1}^{\Lambda + \frac{1}{2}} \sum_{m=\frac{1}{2}}^{\Lambda}
\bigl[
{ { ((m + \zeta)^2 + l (m + l + \zeta))} \over
{ l (m + \zeta)^3 (m + l + \zeta)^4 \  {\cal D}^{(1)}_{m,m+l}(\zeta) } }
\nonumber \\
& {} & \bigl(
 (m + \zeta) (m + l + \zeta) (m - l + \zeta)
- 2 { {((m + \zeta)^2 + l (m + l + \zeta))} \over
    { {\cal D}^{(1)}_{m,m+l}(\zeta) }      }  \bigr) \bigr]
\nonumber \\
& {} & + [\zeta \rightarrow - \zeta] \;.
\nonumber \\
\end{eqnarray}
Evidently, for the high momentum terms in this expression
the fluctuations with respect to $\zeta$ are small due
to the restriction to the
fundamental modular domain and so these contributions
are essentially just constant with respect to $\zeta$. The
dominant contributions to the $\zeta$ dependence are thus
the lowest modes with $m=\frac{1}{2}$. Thus any structure
that can appear is essentially coming from this one degree
of freedom. In fact,
I evaluated the double sum numerically using {\it Mathematica}.
The behaviour of the potential was studied
for different values of the physical dimensionless mass $\rho$.
In addition, there is the parameter $N$ which defines the order
to which the iterations in Eq.(\ref{sumup}) is summed. Curiously,
the convergence as $N\rightarrow\infty$ is very slow,
but stability is achieved by the value $N=10^6$.
As in the part of $V$ independent of $a_0$, the result here has
a logarithmic dependence on the cutoff regulator.
At this point I suppress the divergence by hand, but will
return to the nature of this divergence in the following section.

\begin{figure}
\vspace{1em}
\centerline{
\parbox{1in}{
\vspace*{0.2cm}\psfig{figure=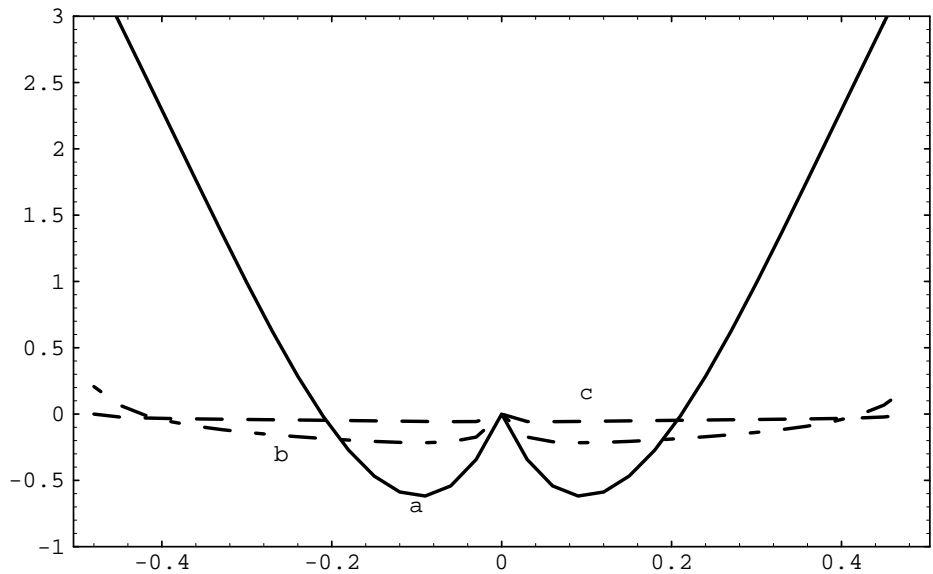,height=17cm}
\vspace*{0.2cm} }
}
\vspace{1em}
\caption{$V_a$ versus $\zeta$, namely
the contribution of $a_0$ to the vacuum potential.
The potential is a functional of $\zeta$ over the fundamental
modular domain $-1/2 < \zeta < 1/2$. The curves are
plotted for various values of renormalised mass $\rho$:
(a) $\rho = 0$ (b) $\rho = 20$ and
(c) $\rho = 100$.  The potential has two minima whose
depth increase with the coupling, namely with decreasing
$\rho$. The wells are at their deepest for effectively
massless `gluons'.}
\label{fig:doublewell}
\end{figure}

{\underline{Shape of the Constrained Part of the Potential.}}
The final result for the renormalised potential, $V_a$, is represented in
Fig.(\ref{fig:doublewell}) for three values of the renormalised
mass $\rho$.
In the deep perturbative region $\rho \rightarrow \infty$
the potential is flat.
However, as the mass $\rho$ is brought down,
or coupling increased, the potential develops two degenerate
minima. The barrier height appears to be at its maximum
precisely at $\rho=0$.
I now estimate its impact.
A least square fit gives parabolae for the shape of
the individual wells. For example, for the
region $0 < \zeta < 0.2$ the curve can be
well described by
\begin{equation}
V_a^{({\rm {fit}})} (\zeta) = 54.9362 \ \zeta^2 - 11.2670 \ \zeta - 0.0385
\;.
\end{equation}
Taking this together with the coefficient of 4 for the kinetic
term in Eq.(\ref{schroed}), one can
estimate the lowest eigenstate if the given well
were the complete potential: $ E_0 \sim 14.8 \gg 0.6$.
The latter number is the barrier height.
Evidently even the lowest states are
too energetic to feel the presence of the barrier.

\section{Open Renormalisation problem}
I now return to the remaining logarithmic divergences
which in the previous calculation were suppressed by hand.
The origin of these divergences is simple to see: in
the vertex $\Gamma_1(\zeta)$ appears the factor
$( { {w_l}\over{u_n} } + { {u_n}\over{w_l} } )$.
This is contracted with the Kronecker delta function
$\delta^n_{m + l}$ in the sums appearing in the
potential. Evidently, for fixed momentum $m$ but $l$ large this
factor scales as a constant
\begin{equation}
\lim_{l\rightarrow\infty} \delta^n_{m + l} \  \bigl(
{ {w_l}\over{u_n} } + { {u_n}\over{w_l} }
\bigr) = 2
\;.
\end{equation}
Consequently, the large $l$ behaviour in the sums
is not suppressed by a contribution from this term.

\begin{figure}
\vspace{1em}
\centerline{
\parbox{1in}{
\vspace*{0.2cm}\psfig{figure=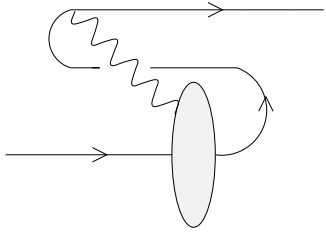,height=3cm}
\vspace*{0.2cm} }}
\vspace{1em}
\caption{Example of a new `Self-Induced Inertia' in the
Hamiltonian, induced
by substituting the constrained zero mode $a_0$ in the
Hamiltonian and bringing the result to normal ordered form.
Such a diagram contains logarithmically divergent coefficients.}
\label{fig:selfinert}
\end{figure}

One eventually realises
this divergence is the tip of the iceberg when one begins
reinserting the solution to the constraint equation
into the Hamiltonian. New many-body operators will be induced,
including two-body operators.
An example is a term arising from a diagram of the
form in Fig.(\ref{fig:selfinert}) with the `R' part of the
$\Gamma_1$ vertex, Eq.(\ref{partvertex}).
The expression corresponding to this diagram is
\begin{equation}
H_{\rm{new}}  \propto
\sum_{m=\frac{1}{2}}^{\Lambda} b^{\dag}_m b_m
\sum_{l=1}^{\Lambda +\frac{1}{2}} \sum_{n=\frac{1}{2}}^{\Lambda}
R^{lmn}(\zeta) C_1^{lmn}(\zeta) \delta^n_{m+l}
\;.
\nonumber
\end{equation}
One observes that the coefficient of the two-point operator
matches the linear contribution of $a_0$ to the adiabatic
potential in Eq.(\ref{linearvaccontr}). The coefficient
logarithmically diverges.
This means that with the
constrained mode eliminated, the two-particle bound state
equation is once again ill-defined and must be renormalised
over again by redefinition of the mass.
There are many new terms contributing
and the process of collecting them up is still underway.
One is faced now with a
subtlety: the constraint equation was rendered finite and
cutoff dependent with the {\it same} renormalisation
used in the two-particle sector. Now that the latter would
appear to have to change, it is not evident how
the renormalisation of the constraint equation could
be analogously modified. The main hurdle is
conceptual rather than merely computational.
It may be that the step of separating the two outstanding
problems in light-cone quantisation -- renormalisation
and the zero modes -- is too naive.

\section{Summary and Conclusions}
I restate the basic motivation for studying this theory: it was
to understand the relationship between dynamical
and constrained zero modes in a non-Abelian gauge theory with some
non-trivial parton spectrum, and whether these modes
contributed to the expected vacuum structure of that theory.
The model was that of two-dimensional SU(2) gauge theory
coupled to massive scalar adjoint fields.
In \cite{PKP95} was shown how a careful gauge-fixing leads to a
linear constraint equation (as also occurs in (3+1) for QCD
\cite{FNP81}) for the zero mode of the real scalar field
in this theory.
This constraint, though at first glance complicated, had a systematic
structure which could be interpreted in terms of propagators
and vertices. A diagrammatic translation of the equation together
with a consideration of the symmetries of the constrained mode
lead to a method of solving the constraint.
I then studied the role of the constrained zero mode
in the effective
potential governing the dynamical or `gauge' zero mode.
It turned out to generate a centrifugal barrier, the potential assuming
a double-oscillator shape. Such a structure is to be expected from
similar pictures in \cite{vBK87}. Evidently,
including the constrained
zero mode makes a qualitative difference in the potential of the
dynamical zero mode in the direction of the
expected result. However, the actual quantitative impact in this theory
appears to be minimal.
The outstanding problem is the completion of the renormalisation,
which here was achieved simply by suppressing the logarithmic
cutoff dependence by hand. A preliminary study of the
further impact of the constrained mode in the
Hamiltonian suggested that these divergences may go hand
in hand with additional singular behaviour in the
new self-induced inertias coming from the
constraint. A detailed treatment of this problem is still
in progress.

The conclusions are three-fold. Firstly, the zero mode has
a structure which can be categorised diagrammatically. This
is immensely useful for solving the relevant constraint equations.
Secondly, when one takes care
to preserve all the modes of the present theory of
massive adjoint scalars coupled to SU(2) glue, there is some
additional structure in the theory originally treated by \cite{DKB93}.
Within the framework of dimensional reduction of
non-Abelian gauge theory, these structures are, at best, suggestive
of vacuum and low-energy phenomena of QCD(3+1).
Thirdly, the problem of renormalisation in the
light-cone approach may not be separable from that of the
zero modes.

\section*{Acknowledgement}
I am grateful to the following for intensive discussions
and insightful suggestions: Hans-Christian Pauli, Steve Pinsky and
Dave Robertson.
I am especially grateful
to Brett van de Sande for many discussions on nearly all aspects of
this work, many insightful suggestions and
patient guidance in the use of {\it Mathematica}. Rolf Bayer is
thanked for assistance in the use of {\it Xfig}.
This work was supported by a Max-Planck Gesellschaft Stipendium.

\begin {appendix}
\setcounter{equation}{0}
\renewcommand{\theequation}{\Alph{section}.\arabic{equation}}
\newpage
\section{Notation and Conventions}
\label{sec:notapp}

\noindent
{\underline {Light-Cone Coordinates.}}
I follow the convention of Kogut and Soper \cite{KoS70},
with $ x^\pm \equiv (x^0 \pm  x^1)/\sqrt{2}$.

\noindent
\underline{Colour Helicity Basis.}
The colour helicity basis for SU(2) is defined in terms of the
Pauli matrices $\sigma^a$:
\begin{equation}
  \tau ^3 = {1\over2} \sigma^3 \ , \quad
  \tau ^\pm \equiv {1\over{2\sqrt{2}}} (\sigma^1 \pm i \sigma^2)
\ .\end{equation}
We can turn this into a vector space by introducing
elements $x^a$ such that tilde quantities are defined
with respect to the helicity basis, and untilded
the usual Cartesian basis:
\begin{equation}
x^a = \left(\begin{array}{c}
		x^1 \\
		x^2 \\
		\end{array}
	\right)
\qquad {\rm {and}}\qquad
\tilde{x}^a = \left(\begin{array}{c}
		\tilde{x}^1 \\
		\tilde{x}^2 \\
		\end{array}
		\right)
	= \left(\begin{array}{c}
		x^+ \\
		x^- \\
	 	\end{array}
	\right).
\end{equation}
The relation between the tilde and untilde basis can be
written
$\tilde{x}^a = \Lambda^a_b x^b
\; {\rm{and}} \;
x^a = \tilde{\Lambda}^a_b \tilde{x}^b $
where $\tilde{\Lambda} = \Lambda^{\dag}$.
With these elements we can construct the metric
in terms of the tilde basis. Essentially we must
demand the invariance of the inner product of
any two vector space elements,
$ x^a y_a = \tilde{x}^a \tilde{y}_a \;.$
Using the fact that the metric in the $a=1,2$
basis is just the Kronecker delta $\delta_{ab}$
and the transformed metric is
$ \tilde{{\cal G}}_{ab} =
\tilde{\Lambda}_a^c \delta_{cd} \tilde{\Lambda}^d_b $.
Thus
\begin{equation}
 \Lambda =  {1\over{\sqrt{2}}}
	\left(\begin{array}{rr}
		1 & i \\
		1 & -i \\
		\end{array}
	\right)
\quad {\rm{and}} \quad
\tilde{{\cal G}}_{ab} =
		\left(	\begin{array}{cc}
			0 & 1 \\
			1 & 0 \\
			\end{array}
		\right) \;.
\end{equation}
The metric to raise and lower indices
in the helicity basis becomes
$ x_{\pm} = x^{\mp} $.
The colour algebra looks formally like
the Lorentzian structure in light-cone coordinates.

\setcounter{equation}{0}
\section {Gribov copies and the Wilson Loop}
\label{sec:GribFock}
\noindent
{\underline{Gribov Copies:}}
Because
of the torus geometry of the space and the non-Abelian
structure of the gauge group, there remain large gauge
transformations which are still symmetries of the theory
\cite{Gri78,Sin78} despite the complete fixing of the theory with
respect to small gauge transformations.
These are generated by local SU(2) elements
\begin{equation}
	U (x^-) = \exp ({- i  n_0 \pi {x^- \over {L}}   \tau _3)},
\;  n_0 \; {\rm{an \; even \;integer}}
\label{gribcopy}
\end{equation}
which satisfy periodic boundary conditions. Another symmetry of the
theory is $Z_2$ centre symmetry which here means allowing for
antiperiodic $V$ or alternately $ n_0$ odd. Both
preserve the periodic boundary conditions on the gauge
potentials.
On the diagonal component of $ A^+$ $U$ generates shifts that
are best expressed in terms of the dimensionless $ z  $,
namely $ z \rightarrow  z ' =  z + n_0$.
On the scalar adjoint field and its momenta
the effect of the transformation is
\begin {eqnarray}
	 \varphi_3  & \rightarrow & \varphi_3 \quad {\rm{and}} \quad
	 \varphi_\pm \rightarrow  \varphi_\pm
		\exp{(\mp i  n_0 {\pi\over L} x^-)},
\label{gribfield} \\
 \pi^3  & \rightarrow & \pi^3 \quad {\rm{and}} \quad
 \pi^\pm \rightarrow \pi^\pm
		\exp{(\pm i  n_0 {\pi\over L} x^-)}.
\label{gribmom}
\end {eqnarray}

\noindent
{\underline{Colour Invariance of $ z $:}}
The gauge mode
$ z $ can be written in terms of a
colour singlet object.
Construct the Wilson line for a contour $ {\rm C} $ along
the $ x $ direction from $- L $ to $ L $
\begin{equation}
 W  = {\rm Tr} {\rm {P}} \exp (i g \int_{{\rm C}} dx_\mu
 {\bf A}^\mu) = {\rm Tr} {\rm {P}}
\exp (i g \int_{- L}^{+ L} d x   {\bf A}^+).
\end{equation}
In the gauge used in this work, this is simply
$  W  = {\rm Tr} \exp (2\, i\, z \,\pi \, \tau ^3)
= 2 \cos (2 \pi  z)
\; .$
Thus, modulo the integers,
$ z  = {1\over{2\pi}} {\rm{arcos}} ({{W}\over2})
\;.$
The integer shifts are the Gribov copies.
Since $ W $ is explicitly constructed in terms of
a colour trace, $ z $ is a colour singlet.

\setcounter{equation}{0}
\section{Commutation Relations for Scalar Fields}
The commutators can be obtained, despite the presence of
the constrained zero mode, from the Dirac procedure
\cite{Dir64}. The result is well known, see for example
\cite{MaY76}, for scalar fields and can be directly taken over
into the present theory.
The quantum commutation relation at equal $ x^+$ for the normal modes is
of the real scalar field is
\begin {equation}
     \Bigl[
{\hbox{\vbox{\ialign{#\crcr
    ${n}$\crcr
   \noalign{\kern1pt\nointerlineskip}
    $\displaystyle{\varphi}_{3}$\crcr}}}}
(x) , \pi^3(y) \Bigr]_{x^+=y^+} =
        {i\over2} \Bigl[ \delta  (x^- - y^-) - {1\over{2 L}} \Bigr] \ ,
\end    {equation}
where the $n$ above the field means the constrained zero mode $a_0$ has been
removed and
the last term ensures consistency for the commutator restricted to
normal mode fields \cite{KaP93}.
For the complex fields they are simply
\begin  {equation}
     \Bigl[ \varphi_-(x) , \pi^-(y) \Bigr]_{x^+=y^+} =
     \Bigl[ \varphi_+(x) , \pi^+(y) \Bigr]_{x^+=y^+} =
        {i\over2} \delta  (x^- - y^-)
\ . \label{fieldcomm} \end    {equation}
These relations demonstrate that there is no subtlety in the
zero modes of the complex fields.

\setcounter{equation}{0}
\section{Fourier transforms, Currents and the Hamiltonian}
\label{sec:charges}
The Discrete Fourier transforms of the scalar current components
are defined by
\begin {equation}
  J_3^+ (x^-) \equiv -{1\over 4 L} \sum_{k \in \ Z}
   e^{-i k {\pi \over  L} x^-} \ {\widetilde J}_3^+ (k)
\ , \qquad {\rm and} \quad
  J_\pm^+ (x^-) \equiv -{1\over 4 L} \sum_{k \in \ H}
   e^{-i k {\pi \over  L} x^-} \ {\widetilde J}_\pm^+ (k)
\ . \label{Fourier}
\end    {equation}
One can verify that
$ \bigl({\widetilde J}^+_3(k) \bigr)^{\dag} =
        {\widetilde J}^+_3(- k) $ and
$ \bigl({\widetilde J}^+_-(k) \bigr)^{\dag} =
        {\widetilde J}^+_+(- k) $.
In the text, as in \cite{PKP95}, the symmetrised operator products
$ B_k \equiv (a_0  b_k + b_k   a_0)/2 $ and
$ D_k \equiv (a_0  d_k + d_k   a_0)/2 $ were introduced.
These carry the $a_0$ dependence in the currents $J_\pm$
leading to the construction of $Q$ operators
\begin {equation}
  {\widetilde J}^+_3(k) \equiv  Q_3 (k) , \quad
   {\widetilde J}^+_+(k) \equiv  Q_+ (k)
     + {D_k \over  v_k}       , \quad\  {\rm and}\quad
     -{\widetilde J}^+_-(k) \equiv  Q_+ (k)
     + {B_k \over  u_k}
\ .
\label{defcharge}
\end    {equation}
The $Q$-operators are thus $a_0$ independent.
It suffices here to give the charge operator $Q_-$ explicitly
\begin {eqnarray}
   Q_- (k) =  \sum_{n =  1}^\infty
                 \sum_{m = {1\over2}}^\infty
        a_n  b_m \ \Bigl({w_n \over  u_m}
                            - {u_m \over  w_n} \Bigr)
                     \delta^k_{n + m}
           & + & \sum_{n =  1}^\infty
               \sum_{m = {1\over2}}^\infty
        a^{\dag}_n  b_m \ \Bigl({w_n \over  u_m}
                            + {u_m \over  w_n} \Bigr)
                     \delta^m_{n + k}
\nonumber \\
           & - & \sum_{n = 1}^\infty
               \sum_{m = {1\over2}}^\infty
        a_n  d^{\dag}_m \ \Bigl({w_n \over  v_m}
                            + {v_m \over  w_n} \Bigr)
                     \delta^n_{m + k}
\label{charge-}
\end   {eqnarray}
with $Q_+ = (Q_-)^{\dag}$.

The Hamiltonian is obtained from the energy-momentum
tensor $  \Theta^{\mu\nu} = 2{\rm Tr} \bigl({\bf F}^{\mu\kappa}
   {\bf F}_\kappa^{\phantom{\kappa}\nu}\bigr)
 - g^{\mu\nu} {\cal L}  $.
At one level it is a simple expression, standard for
(1+1)-dimensional gauge theories.
\begin {eqnarray}
    P^- & = & \int_{- L}^{+ L} \!\! dx^- {\rm Tr}
 \ \bigl(\partial_+ {\bf A}^+ - {\bf D}_- {\bf  A}^- \bigr)^2
           = \int_{- L}^{+ L} \!\! dx^- {\rm Tr}
 \ \bigl(\ \partial_+ {\bf A}^+ \partial_+ {\bf A}^+
  - g^2  {\bf J}^+{1\over {\bf D}_-^2} {\bf J}^+
 \ \bigr)
\nonumber \\
\end{eqnarray}
to which must be added the mass term
$ \mu_0^2 \int dx^- {\rm{Tr}}{\bf \Phi}^2 $.
The above form merely expresses the dominance of the Coulomb potential
in (1+1) dimensions. However unpacking the currents $J^+$
leads to nontrivial structure.
The length dimension can be factored out by defining
$H$ via $ P^- =  L \Bigl(g  / 4\pi\Bigr)^2  H $.
In terms of Fourier components, the gauge mode and
Coulomb parts of $H$ can be rewritten as
\begin  {eqnarray}
 H  =
 - 4{d^2 \over d z^2}
 + \sum_{k =1}^{\infty}
     w_k^4 \bigl(\widetilde J_3^{\dag} (k) \widetilde J_3 (k)
         + \widetilde J_3 (k) \widetilde J_3^{\dag} (k) \bigr)
 & + & \sum_{k ={1\over2}}^{\infty}
     v_k^4 \bigl(\widetilde J_+^{\dag} (k) \widetilde J_+ (k)
         + \widetilde J_+ (k) \widetilde J_+^{\dag} (k) \bigr)
\nonumber \\
 & + & \sum_{k ={1\over2}}^{\infty}
     u_k^4 \bigl(\widetilde J_-^{\dag} (k) \widetilde J_- (k)
         + \widetilde J_- (k) \widetilde J_-^{\dag} (k) \bigr)
  \ .
\nonumber \\
\label{hamall}
\end   {eqnarray}
Relating $J$ operators to the $Q$ operators one can
separate the constrained mode $a_0$ parts of H from the
Fock sector.
The reader is referred to \cite{PKP95} for more detail.
When combined into the invariant mass-squared operator
$M^2 = 2P^+ P^-$, with $P^+$ the momentum operator, $L$
completely drops out in favour of
$K = (L/\pi) P^+$, the harmonic resolution.
The continuum limit $L\rightarrow \infty$ translates into
$K \rightarrow \infty$ which is easily implementable
in computer simulations.

\end{appendix}


\newpage
\begin {thebibliography}{30}
\bibitem {Wei66}
        S.~Weinberg, {\it Phys.Rev.} {\bf 150} (1966) 1313.
\bibitem {LeB80}
  G.P.~Lepage and S.J.~Brodsky,
     {\it Phys.Rev.} {\bf D22} (1980) 2157.
\bibitem {PaB85}
  H.C.~Pauli and S.J.~Brodsky,
     {\it Phys.Rev.} {\bf D32} (1985) 1993; 2001.
\bibitem {PHW90}
   R.J.~Perry, A.~Harindranath and K.~Wilson,
      {\it Phys.Rev.Lett.} {\bf 65} (1990) 2959.
\bibitem {Dir49}
  P.A.M. Dirac, {\it Rev.Mod.Phys.} {\bf 21} (1949) 392.
\bibitem {HeK95}
  M. Heyssler, A.C. Kalloniatis,
  {\it Phys.Lett.} {\bf B 354} (1995) 453.
\bibitem {Rob93}
        D.G.~Robertson, {\it Phys.Rev.} {\bf D47} (1993) 2549.
\bibitem {MaY76}
   T.~Maskawa and K.~Yamawaki,
   {\it Prog.Theor.Phys.} {\bf 56} (1976) 270.
\bibitem {HKW91}
  T.~Heinzl, S.~Krusche, E.~Werner and B.~Zellermann,
     {\it Phys.Lett.} {\bf B272} (1991) 54,
\bibitem {BPV93}
    C.M.~Bender, S.S.~Pinsky, B.~van~de~Sande,
    {\it Phys.Rev.} {\bf D48} (1993) 816.
\bibitem {PiV94}
    S.S.~Pinsky and B.~van~de~Sande,
    {\it Phys.Rev.} {\bf D49} (1994) 2001.
\bibitem {PVH95}
    S.S.~Pinsky, B.~van~de~Sande and J.~Hiller,
    {\it Phys.Rev.} {\bf D51} (1995) 726.
\bibitem{BGW95}
    A. Borderies, P. Grang\'e, E. Werner,
    {\it Phys.Lett.} {\bf B 345} (1995) 458.
\bibitem {SVZ79}
  M.A. Shifman, A.I. Vainshtein and V.I. Zakharov,
  {\it Nucl. Phys.} {\bf B147} (1979) 385.
\bibitem {FNP81}
   V.A.~Franke, Yu.A.~Novozhilov, and E.V.~Prokhvatilov,
       {\it Lett.Math.Phys.} {\bf 5} (1981) 239;
   437;  
      {\it Light-Cone Quantization of gauge Theories with
           periodic Boundary Conditions}
       in {\it Dynamical Systems and Microphysics},
       Academic Press, 1982, p.389-400.
\bibitem {DKB93}
   K.~Demeterfi, I.R.~Klebanov, and G.~Bhanot,
   {\it Nucl.Phys.} {\bf B418} (1994), 15; 
   G.~Bhanot, K.~Demeterfi, and I.R.~Klebanov,
   {\it Phys.Rev.} {\bf D48} (1993) 4980. 
\bibitem {PKP95}
   H.C. Pauli, A.C. Kalloniatis, S.S.Pinsky,
   {\it Towards Solving QCD -- The Transverse Zero Modes in
   Light-Cone Quantisation}. hep-th/9509020.
   To appear in {\it Phys. Rev.} D;
   S.S. Pinsky, A.C. Kalloniatis,
   {\it Light-Front QCD(1+1) Coupled to Adjoint Scalar Matter}.
   Submitted to {\it Phys. Lett.} B.
\bibitem {PaB95}
   H.-C. Pauli, R. Bayer,
   {\it Towards Solving QCD in Light-Cone Quantization --
   On the Spectrum of the Transverse Zero Modes for SU(2)}
   Heidelberg Preprint MPI H-V32-1995.
\bibitem {Man85}
   N.S.~Manton,
   {\it Ann.Phys.(N.Y.)} {\bf 159} (1985) 220.
\bibitem{LST94}
   F.~Lenz, M.~Shifman, M.~Thies, 1994.
\bibitem{LNT94} F. Lenz, H.W.L. Naus, M. Thies,
  {\it Ann. Phys. (N.Y.)} {\bf 233} (1994) 317.
\bibitem {KaP94}
  A.C.~Kalloniatis and H.C.~Pauli,
        {\it Z.Phys.} {\bf C63} (1994) 161.
\bibitem {KaR94}
  A.C.~Kalloniatis and D.G.~Robertson,
  {\it Phys.Rev.} {\bf D50} (1994) 5262. 
\bibitem {KPP94}
  A.C.~Kalloniatis, H.C.~Pauli, and S.S.~Pinsky,
   {\it Phys.Rev.} {\bf D50} (1994) 6633. 
\bibitem{Lus83}
   M. L\"uscher,
    {\it Nucl.Phys.} {\bf B219} (1983) 233.
\bibitem {vBa92}
   P. van Baal,
   {\it Nucl.Phys.} {\bf B369} (1992) 259.
\bibitem{Kal94}
   A.~C.~Kalloniatis,
   {\it Solving the Zero Mode Problem of QCD},
   in Proceedings of Workshop on Hadrons and Light-Front QCD,
   Zakopane, Ed. St. Glazek, 1994.
\bibitem {Het93}
   J.E.~Hetrick,
   {\it Nucl.Phys.} {\bf B30} (1993) 228;
        {\it Int.J.Mod. Phys.} {\bf A9} (1994) 3153.
\bibitem{vBK87}
   P. van Baal, J. Koller,
   {\it Ann.Phys.(N.Y.)} {\bf 174} (1987) 299.
\bibitem {KoS70}
   J.B. Kogut, D.E. Soper,
     {\it Phys.Rev.} {\bf D1} (1970) 2901.
\bibitem{Gri78}
   V.N. Gribov,
   {\it Nucl.Phys.} {\bf B} 139 (1978) 1.
\bibitem{Sin78}
   I.M. Singer,
   {\it Commun.Math.Phys.} {\bf 60} (1978) 7.
\bibitem{Dal95}
   S. Dalley, private communication.
\bibitem {Dir64}
   P.A.M.~Dirac,
   {\it Lectures on Quantum Mechanics.}
   (Academic Press, New York, 1964)
\bibitem{KaP93}
   A.~C.~Kalloniatis, H.-C.~Pauli,
  {\it Z.Phys.} {\bf C60} (1993) 255.
\end {thebibliography}
 \end   {document}